\def\ni {\noindent}

\def\ssk{\vskip  5truept}

\def\bsk{\vskip 15truept}
\def\newpage{\vfill\eject}
\def\newline{\hfil\break}
\def\reff{\par\noindent\hangindent 15pt}

\def\gx31{GX3\raisebox{1pt}{+}1}
\def\sun{\hbox{$\odot$}}
\def\arcmin{\hbox{$^\prime$}}

\def\fdg{\hbox{$.\!\!^\circ$}}
\documentstyle[epsfig]{article}

\begin{document}

\voffset 0truein
\hsize 5truein
\vsize 8truein

\font\abstract=cmr8
\font\keywords=cmr8
\font\caption=cmr8
\font\references=cmr8
\font\text=cmr10
\font\affiliation=cmssi10
\font\author=cmss10
\font\mc=cmss8
\font\title=cmssbx10 scaled\magstep2
\font\titlep=cmssbx10 scaled\magstep1
\font\alcit=cmti7 scaled\magstephalf
\font\alcin=cmr6 
\font\ita=cmti8
\font\mma=cmr8

\null

\title{\ni GRANAT$\!$/$\!$ART{\titlep \raisebox{2pt}{-}}P OBSERVATIONS
OF\,GX3{\titlep \raisebox{2pt}{+}}1: $\!$\mbox{TYPE\,I} X{\titlep
\raisebox{2pt}{-}}RAY BURST AND PERSISTENT EMISSION} 

\bsk \bsk

%\vspace{-0.3cm}
\author{\ni S.V.Molkov$^{\,1,2}$, S.A.Grebenev$^{\,1,2}$, M.N.Pavlinsky$^{\,1,2}$, R.A.Sunyaev$^{\,1,2}$}
\bsk
\affiliation{1) Space Research Institute, Russian Academy of Sciences,
Moscow, Russia\\ \indent 2) Max-Planck-Institut f\"{u}r
Astrophysik, Garching, Germany}                                                

\bsk
\baselineskip = 12pt

\vspace{-1mm}
\abstract{ABSTRACT \ni 
We present results of observations of the known LMXB source GX3+1 with
the telescope ART-P on board GRANAT in the fall of 1990. A strong X-ray
burst was detected from the source on Oct. 14\, when it was in the low
X-ray state with a luminosity {\footnotesize $\sim 30$\%}\, smaller than
the normal one. That was only the second case for the whole history of
its study when it exhibited such type of activity. We describe results
of the source spectroscopy during the burst and persistent state and
discuss formation of its X-ray emission in the accretion disk boundary
layer. It is noted that scattering on electrons plays the dominant role
in the emission processes.}

\vspace{-1mm}
\bsk
\baselineskip = 12pt

\bsk
\baselineskip = 12pt
\vspace{-1mm}

\text{\ni 1. INTRODUCTION
\ssk
\ni     
The source \gx31\ is a typical representative of the group of
bright Galactic \mbox{X-ray} binaries consisting of a neutron star with
weak magnetic field and a low mass ($M_1\leq M_{\sun}$)
companion. Sources of this group radiate due to disk accretion and
the bulk of the observed X-rays originates from the geometrically thin
plasma layer at the boundary between the disk inner edge and the
neutron star surface.

%--------------------------- figure 1 --------------------------------
\begin{figure}[pt]
\vspace{-2.5mm}
\mbox{
%\begin{minipage}[b]{57mm}
%\hspace{0.5mm}\psfig{file=fig_lcurve.eps, width=51mm}
%\end{minipage} \hspace{0.2mm}\begin{minipage}[b]{67mm}
%\psfig{file=fig_burst.ps, width=63.5mm}
%\end{minipage}
\psfig{file=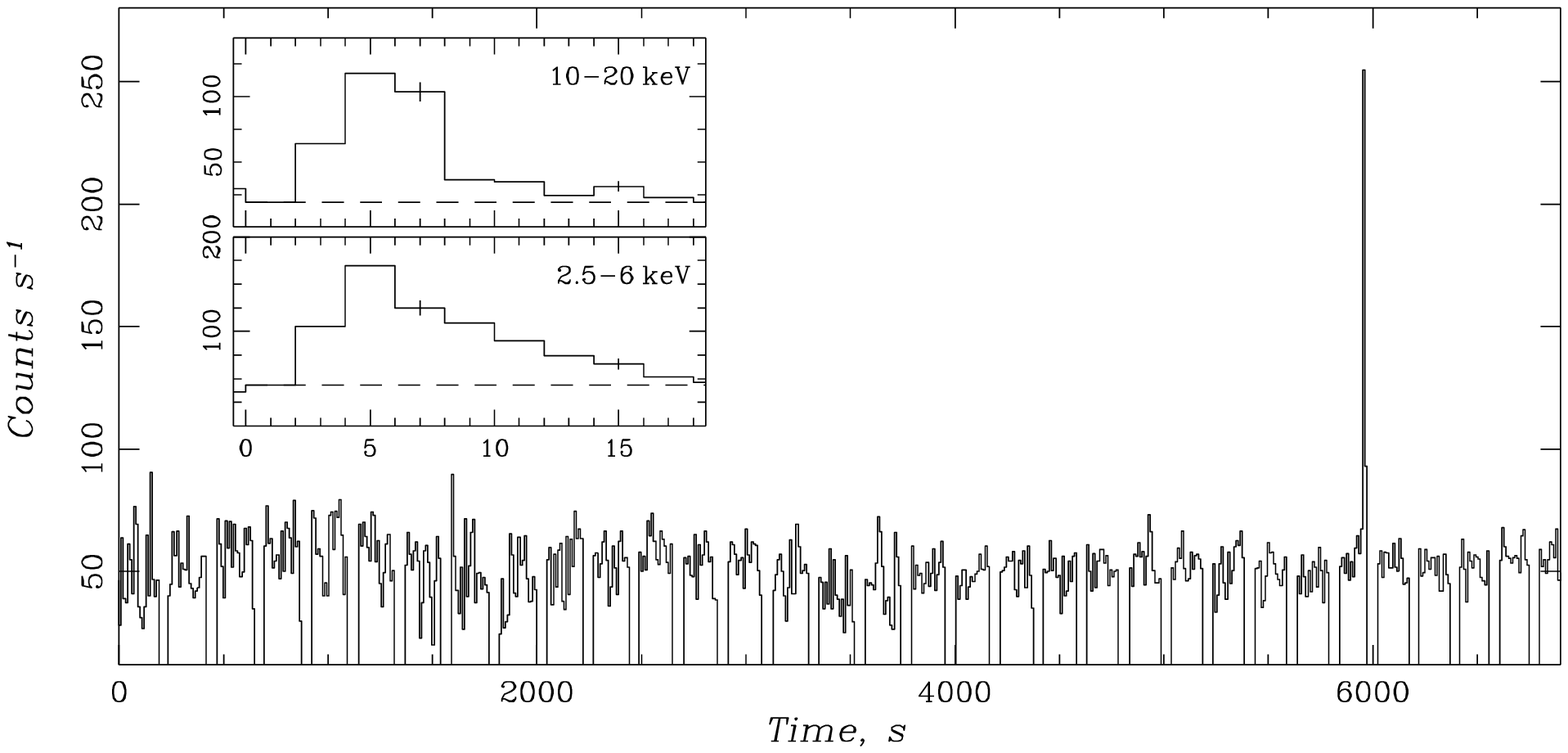, width=120mm}}\\

\vspace{-2.4mm}
\makebox[1ex]{}\begin{minipage}{4.8truein}
\caption{FIGURE 1. The 2.5-20 keV count rate history recorded by ART-P 
on Oct.~14, 1990 when a strong X-ray burst was detected from
\gx31. The insertion gives the burst profile in two different energy
bands with the better temporal resolution (bars show {\footnotesize
$\pm 1\,\sigma$} errors).} 
\end{minipage}

%--------------------------- figure 2 --------------------------------
\vspace{2mm}
\centerline{\hspace{6mm}\psfig{file=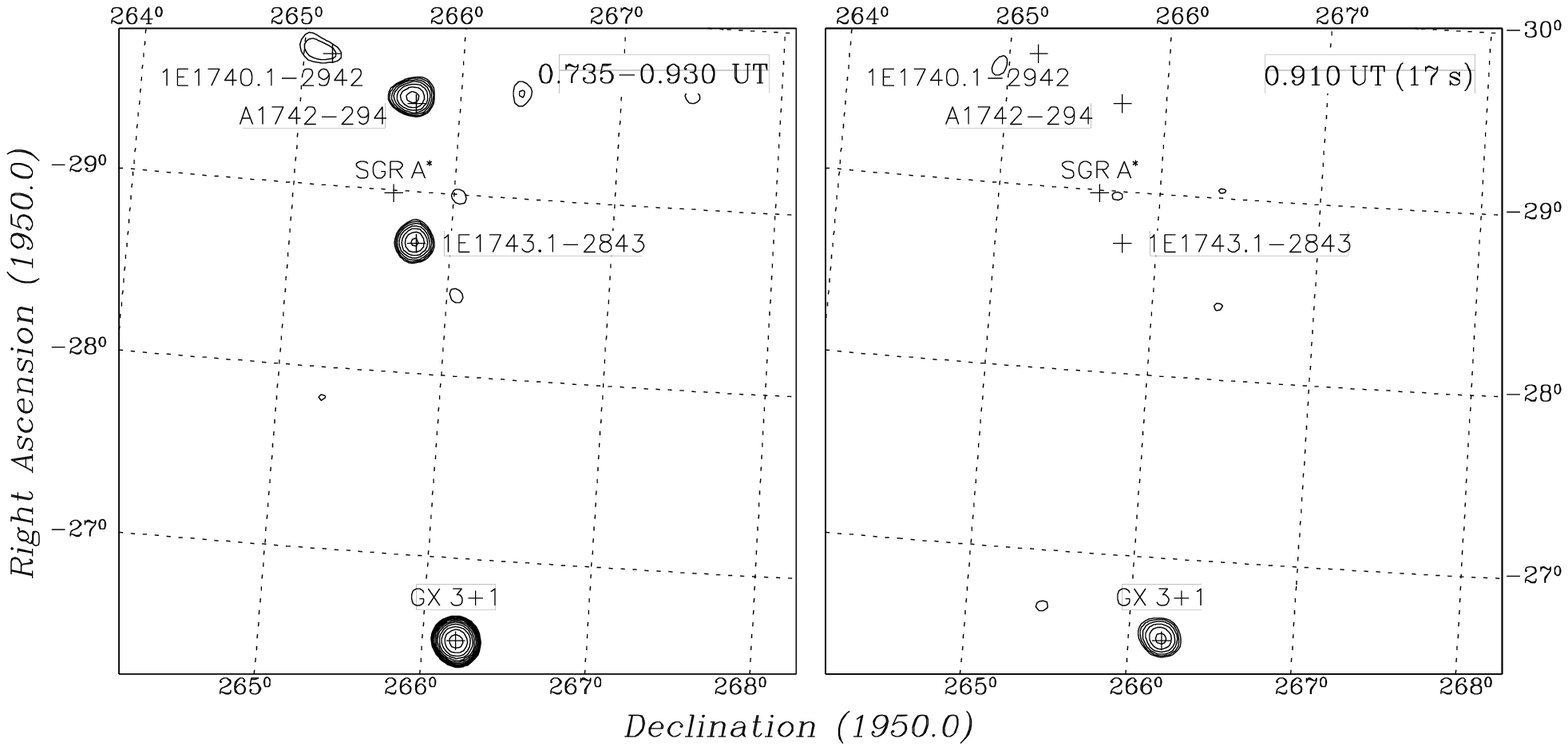, width=4.8truein}}

\vspace{3.7mm}
\makebox[1ex]{}\begin{minipage}{4.8truein}
\caption{FIGURE 2. Image of the field near \gx31\ obtained with ART-P
in the 2.5-20 keV X-rays during the whole observation on Oct.~14, 1990
(left panel) and during the burst interval (right panel). Contours are
given at the signal-to-noise levels of 3, 3.9, 5.0, 6.6, 8.6, 11.1,
..., 53.8\,{\footnotesize $\sigma$}.}
\end{minipage}

\vspace{4mm}
{\begin{minipage}{5truein}
\text
\ni 
the source responsible for the detected burst event. The
insertion in Fig.\,1 shows that the burst duration $\Delta t$
noticeably changed with energy being of about 12 s in the hard 10-20
keV band and reaching $\sim18$ s in the soft 2.5-6 keV band. Such
behaviour is a characteristic of the type I X-ray bursts (Hoffman et
al. 1978).

\hspace{3ex} The source persistent luminosity $L_{p}$ was equal to
$(5.4\pm0.2)\times10^{37}\ \mbox{erg s}^{-1}$ in the 2.5-20 keV band
(assuming a distance of 8.5 kpc). This is $\sim 30$\% less than the
luminosity measured in the other days. The luminosity during the burst
interval $L_{b}\simeq3\times10^{38}\, \mbox{erg s}^{-1}.$ The black-body
approximation of the burst average spectrum gave the neutron star radius
$R\simeq 7.2\pm1.2$ km and its surface temperature $kT_b\simeq2.4\pm0.2$
keV. \ The burst recurrence time was estimated to be \ $t_r\sim \Delta
t\, \nu\, L_{b}/L_p\simeq$ 
\end{minipage}}
\end{figure}
%---------------------------------------------------------------------

The coded-mask X-ray telescope ART-P, one of two major instruments on
board GRANAT, was designed for imaging, timing and spectroscopy of
compact sources in the 2.5-60 keV band. It has good angular (5\arcmin)
and temporal (3.9 ms) but moderate energy (22\% at 6 keV) resolution
(Sunyaev et al.\,1990). \gx31\ was observed with ART-P four times
during the GRANAT Galactic center field survey in the fall of 1990. Each of
the observations lasted several hours and the total exposure exceeded
18 hours. Already the quick-look analysis of the data led to
one interesting finding -- a strong X-ray burst was detected from \gx31\
on Oct.\,14 (Pavlinsky et al.\,1994).

\vspace{-1mm}
\bsk
\ni 2. X-RAY BURST
\ssk
\ni 
The 2.5-20 keV count rate recorded by ART-P on Oct.\,14, 1990 is shown as a
function of time in Fig.\,1. The burst occurred $\sim6000$ s
after the session beginning (at $21^h49^m59^s$ UT). In addition to
\gx31\ there were three other persistent X-ray sources present that day
within the telescope's $3\fdg4\times3\fdg6$ field of view (Fig.\,2,\,
left). One of them was the X-ray burster A1742-294. The \mbox{ART-P}
image accumulated during a 17 s burst interval (Fig.~2, right)
evidently shows that only \gx31\ can be 
\newpage
%--------------------------- figure 3 ----------------------------------
\begin{figure}[h]
\vspace{-3.5mm}
\mbox{
\begin{minipage}[b]{55mm}
\hspace{-3mm}\psfig{file=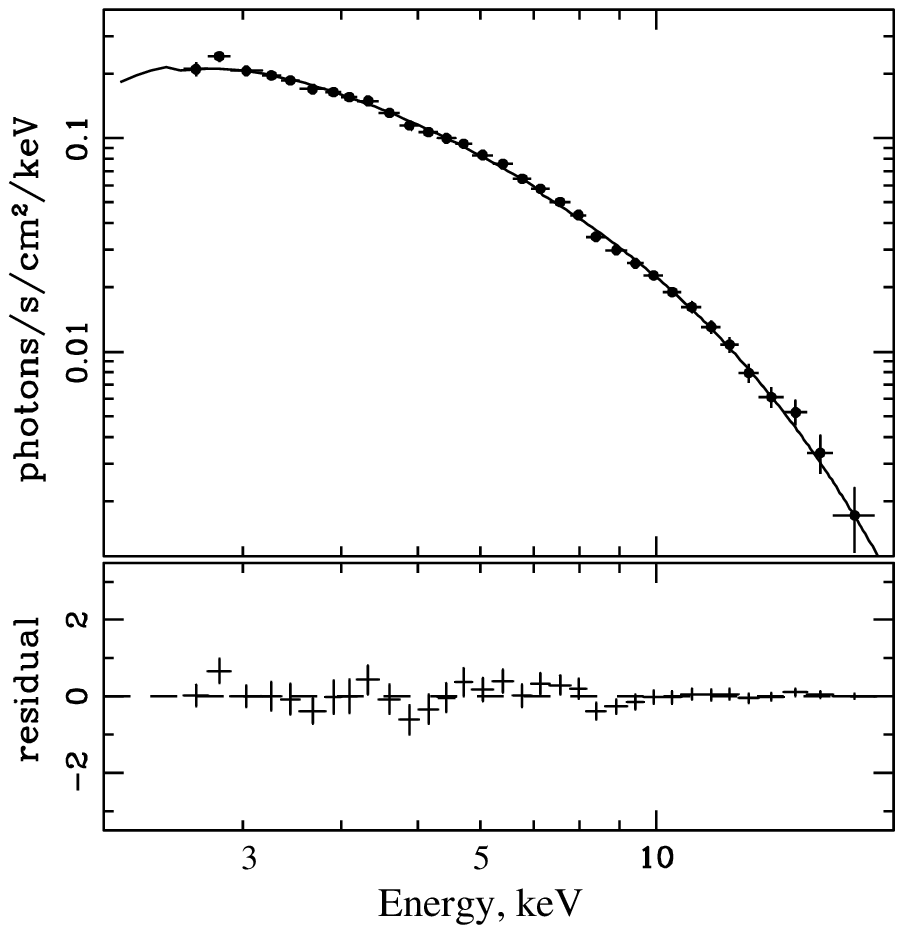, width=54mm}
\end{minipage} \hspace{-3mm}\begin{minipage}[b]{73.5mm}
\psfig{file=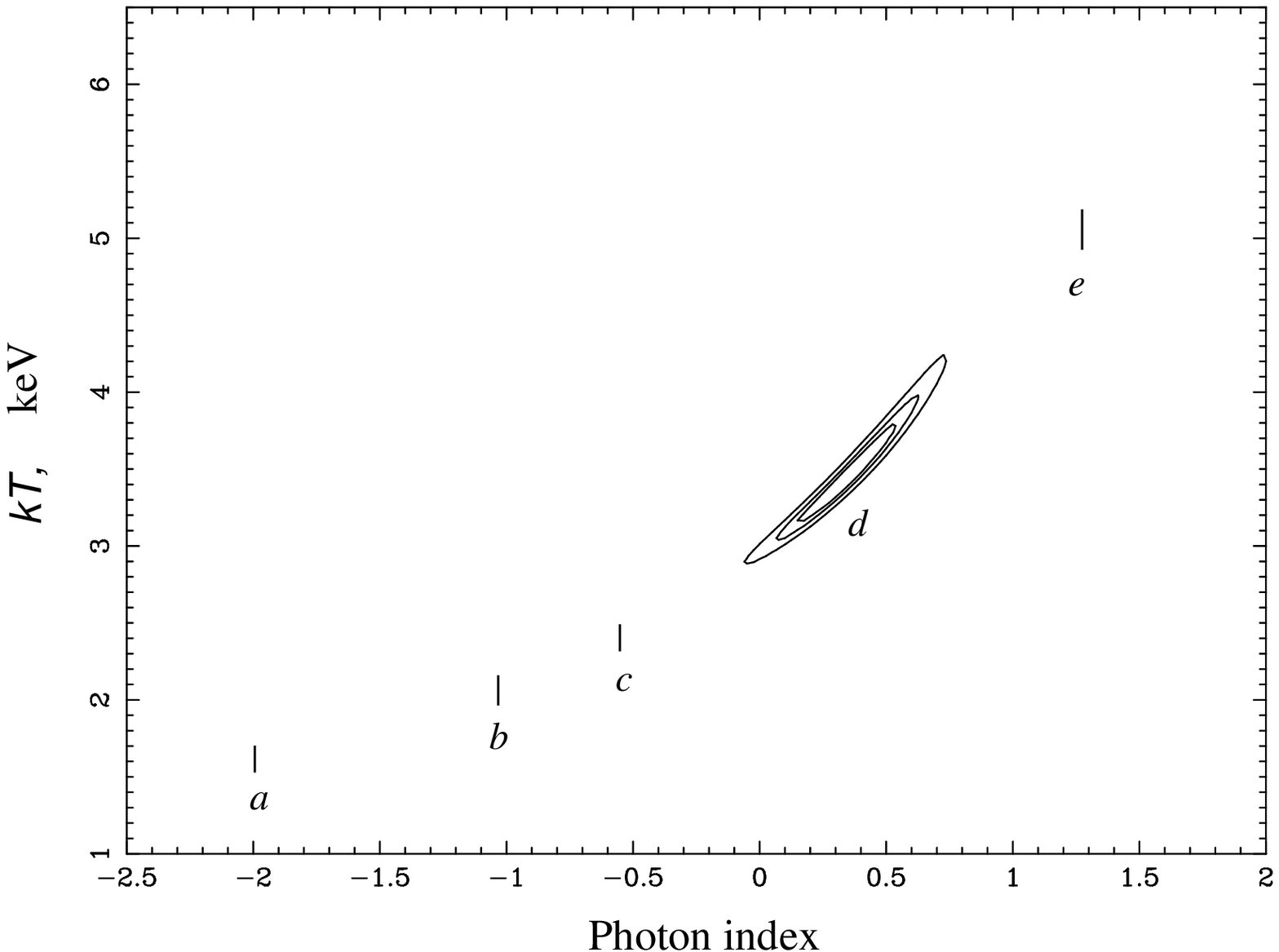, width=72.7mm}
\end{minipage}} \\

\vspace{-4.8mm}
\makebox[1ex]{}\begin{minipage}{4.8truein}
\caption{FIGURE 3. The \gx31\  spectrum  measured with ART-P
on Sept.\,13 and its best-fit approximation by Sunyaev-Titarchuk
Comptonization model (left panel). Right panel shows results of the
spectrum approximation by other models. Given are the {\footnotesize $kT
-\alpha$}\ confidence intervals (1,~2, 3\,{\footnotesize $\sigma$} for
model {\footnotesize\it d\/} and 3\,{\footnotesize $\sigma$} for others)
for black-body ({\footnotesize\it a\/}), exponential
atmosphere\,({\footnotesize\it b\/}), uniform isothermal
half-space\,({\footnotesize\it c\/}), Boltzmann law ({\footnotesize\it
d\/}) and thermal bremsstrahlung ({\footnotesize\it e\/}) models.}
\end{minipage}

\vspace{-0.6mm}
\end{figure}
%---------------------------------------------------------------------

\ni
$10^4$ s, where $\nu\simeq100$ is the ratio of
energy emitted during the accretion onto a neutron star and that
released in thermonuclear reactions of helium burning. This time
exceeds twice the duration of the source observation with ART-P on
Oct.\,14.

Although \gx31\ is one of the brightest X-ray sources on the sky,
located near the Galactic center and thus regularly observed, that was
only the second case for the whole history of its study when it
exhibited bursting activity. First several bursts were detected by
HAKUCHO in 1980 (Makishima et al. 1983). In both the cases the source
was in the low X-ray state with the luminosity 30--50\% smaller than
the normal one. This supports the idea that bursting activity of
many LMXB sources is suppressed due to their high accretion rates
(Lewin et al. 1993). 
%--------------------------- figure 4 ----------------------------------
\begin{figure}
\vspace{-2.5mm}
\centerline{\psfig{file=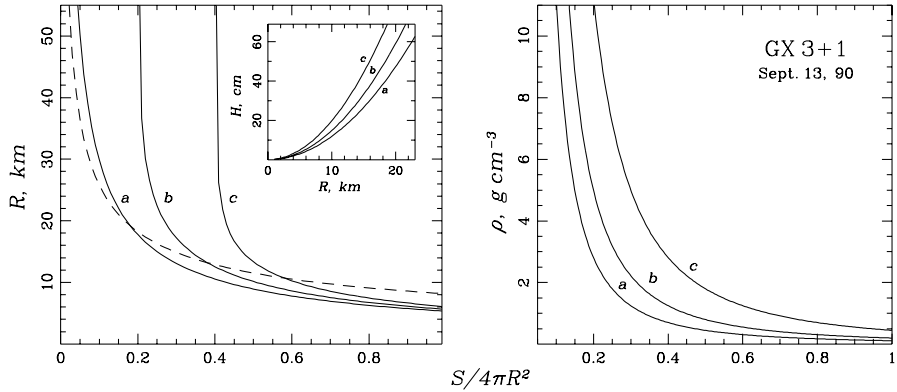, width=12cm}}

\vspace{0.5mm}

\makebox[1ex]{}\begin{minipage}{4.8truein}
\caption{FIGURE 4. Radius {\footnotesize $R$} of a neutron star in \gx31 and 
parameters of its atmosphere, height {\footnotesize $h$} and density
{\footnotesize $\rho$}, estimated from the ART-P data of Sept.\,13\, on
the basis of models of exponential atmosphere (left) and uniform
isothermal half-space (right) and given as functions of the neutron star
emitting surface area {\footnotesize $S$}. In the left panel letters
{\footnotesize ({\it a\/}, {\it b\/}, {\it c\/})} mark cases with
different luminosity levels {\footnotesize $L\ll L_{\rm cr},\ L=0.2\
\mbox{or}\ 0.4\, L_{\rm cr}$}, in the right panel -- with different
suggestions about the neutron star radius {\footnotesize $R=4,\ 3\
\mbox{or}\ 2\, R_{\rm g}$}. Here {\footnotesize $L_{\rm cr}\simeq
1.8\times10^{38}\ \mbox{erg s}^{-1}$ and $R_{\rm g}\simeq 4.2$} km are
the critical Eddington luminosity and the gravitational radius of a
{\footnotesize $1.4\ M_{\sun}$} neutron star. Dashed line shows the
dependence {\footnotesize $R$ on $S$} for the black-body emission model.}
\end{minipage}
\vspace{-3.5mm}
\end{figure}
%---------------------------------------------------------------------

\vspace{-1.mm}
\bsk
\ni 3. PERSISTENT EMISSION AND PARAMETERS OF THE NEUTRON STAR
\ssk
\ni 
The accretion disk boundary layer gives the main contribution to the
\gx31\ persistent emission, so the source spectra measured with ART-P
can be used to investigate spectral formation in it. First, we
applied to the spectra a few simple models, such as black-body
emission, optically thin thermal bremsstrahlung, Sunyaev-Titarchuk
Comptonization of soft photons on hot electrons and Boltzmann law in
which photon flux
$I_{\nu}\sim E^{-\alpha}\,\exp{(-E/kT)}$. The Comptonization model was
most successful. The left panel in Fig.~3 illustrates this issue
showing the spectrum measured on Sept.\,13, 1990 (during the longest
ART-P observation) and the result of its best-fit approximation.\,The
obtained plasma cloud parameters, $kT\simeq 2.3\pm 0.1$\,keV and $\tau_{\rm
T}\simeq14\pm1$, indicate however that the model may be physically
invalid because it does not take into account effects of free-free
absorption important in such a cool and optically thick plasma. Two
other models more suitable for the description of boundary layer
emission were considered -- the model of exponential isothermal
atmosphere and the model of uniform isothermal half-space. They take
into account free-free absorption but still assume electron scattering
to be dominant in opacity. In the first model $I_{\nu}\sim
h^{-{1}/{3}}\,E^{\,1}\,\exp{(-{E}/{kT})}$, in the second $I_{\rm
\nu}\sim\sqrt{\rho}\,E^{{1}/{2}}\,\exp{(-{E}/{kT})}$ (Shakura,
Sunyaev 1973). Here $h$ and $\rho$ are the atmosphere scale height and
its density.  

The models approximate the data quite successfully. In
addition, they provide us with rather interesting estimates for the
neutron star radius and the atmosphere parameters $h$ and $\rho$
(Fig.\,4, see Molkov et al. 1999 for more details). Noticing that these
two and some others of the considered models have the same spectral
shape (Boltzmann law) we presented in Fig.~3 (right panel) confidence
intervals for best-fit parameters (photon index $\alpha$
and $kT$) of the models. The figure shows that the density distribution in the
boundary layer is uniform rather than exponential\,and that the finite
thickness of the layer can not be neglected in modeling of its
spectrum.}

\vspace{-2mm}
\bsk
\baselineskip = 12pt
{\abstract \ni ACKNOWLEDGMENTS

\ni This work was supported by RBRF grants 96-02-18212, 98-02-17056
and INTAS grant 93-3364-ext
}

\vspace{-2mm}
\bsk
\baselineskip = 12pt
{\references \ni REFERENCES
\ssk

\vspace{-1.3mm}
\reff Hoffman, J.A., Marshall, H.L., Lewin, W.H.G. 1978, Nature, 271, 630

\reff Lewin, W.H.G., Van Paradijs, J., Taam, R.E. 1993, Space Science
     Rev., 62, 223

\reff Makishima, K., Mitsuda, K., Inoue, H., et al. 1983, ApJ, 267, 310

\reff Molkov, S.V., Grebenev, S.A., Pavlinsky, M.N., Sunyaev,
R.A. 1999, Astron. Lett., in press

\reff Pavlinsky, M.N., Grebenev, S.A., Sunyaev, R.A. 1994, ApJ, 425, 110  

\reff Shakura, N.I., Sunyaev, R.A. 1973, A\&A, 24, 337

%\reff Sunyaev, R.A., Titarchuk, L.G. 1980, A\&A, 86, 121

%\reff Shakura, N.I. 1972, Sov. Astron., 16, 532
}                      
\end{document}